\documentclass{article}
\usepackage[utf8]{inputenc}
\usepackage[T1]{fontenc}
\usepackage{geometry}
\usepackage{amsfonts}
\usepackage{spconf}
\usepackage{amsmath}
\usepackage{graphicx}
\usepackage[hyphens]{url}
\usepackage{array}
\usepackage{float}
\usepackage{caption}
\usepackage[space]{grffile}
\usepackage{stmaryrd}
\usepackage{subfig}
\usepackage[section]{placeins}
\usepackage{algpseudocode}
\usepackage{color}
\usepackage{enumitem}
\usepackage{times}
\usepackage[makeroom]{cancel}
\usepackage{MnSymbol}
\usepackage{longtable}

\usepackage{etoolbox}
\apptocmd{\thebibliography}{\footnotesize}{}{}
\usepackage{balance}

\DeclareMathOperator{\trace}{tr}

\newcommand{\E}{\mathbb{E}}
\newcommand{\mydef}{:=}
\def\mbf#1{\mathbf{#1}}
\def\mbb#1{\mathbb{#1}}
\def\bs#1{\boldsymbol{#1}}

\def\myparagraph#1{\smallskip\smallskip\noindent\textbf{#1. \hspace{.1cm}}}

\makeatletter
\newcounter{savesection}
\newcounter{apdxsection}
\renewcommand\appendix{\par
  \setcounter{savesection}{\value{section}}%
  \setcounter{section}{\value{apdxsection}}%
  \setcounter{subsection}{0}%
  \gdef\thesection{\@Alph\c@section}}
\newcommand\unappendix{\par
  \setcounter{apdxsection}{\value{section}}%
  \setcounter{section}{\value{savesection}}%
  \setcounter{subsection}{0}%
  \gdef\thesection{\@arabic\c@section}}
\makeatother

\setlist[itemize]{label=$\triangleright$}

\usepackage[hang,flushmargin]{footmisc}

\title{Deep learning-based probabilistic speech model for semi-supervised multichannel speech enhancement}
\title{Semi-supervised multichannel speech enhancement with \\deep learning-based probabilistic speech modeling}
\title{Semi-supervised multichannel speech enhancement with\\ variational autoencoders and non-negative matrix factorization}

\name{Simon Leglaive\textsuperscript{\normalfont 1}\thanks{This work is supported by the ERC Advanced Grant VHIA \#340113.} \qquad Laurent Girin\textsuperscript{\normalfont 1,2} \qquad Radu Horaud\textsuperscript{\normalfont 1}}

\address{\textsuperscript{1}Inria Grenoble Rh\^one-Alpes, France \qquad \textsuperscript{2}Univ. Grenoble Alpes, Grenoble INP, GIPSA-lab, France}

\begin{document}
\ninept
\maketitle

\begin{abstract}
In this paper we address speaker-independent multichannel speech enhancement in unknown noisy environments. Our work is based on a well-established multichannel local Gaussian modeling framework. We propose to use a neural network for modeling the speech spectro-temporal content. The parameters of this supervised model are learned using the framework of variational autoencoders. The noisy recording environment is supposed to be unknown, so the noise spectro-temporal modeling remains unsupervised and is based on non-negative matrix factorization (NMF). We develop a Monte Carlo expectation-maximization algorithm and we experimentally show that the proposed approach outperforms its NMF-based counterpart, where speech is modeled using supervised NMF.
\end{abstract}
\section{Introduction}
\label{sec:intro}

Speech enhancement is a classical problem of speech processing which aims at recovering a clean speech signal from a noisy recording \cite{loizou2007speech}. In this work we focus on multichannel speech enhancement in additive noise, with static (non moving) speech and noise sources. We follow a statistical approach based on a local Gaussian model of the time-frequency (complex-valued) signal coefficients. This approach has been very popular for addressing the audio source separation problem  \cite{ozerov2010multichannel,duong2010under,arberet2010nonnegative,ozerov2012general,sawada2013multichannel,kitamura2016determined,nugraha_taslp16,leglaive:taslp16}. 
In the multichannel case, the covariance matrix of this model can be structured as the product of a time-frequency-dependent \emph{variance}, accounting for the spectro-temporal content of the source signal, and a frequency-dependent \emph{spatial covariance matrix} (SCM), accounting for the spatial properties of the image signal \cite{duong2010under}. The variance term was further structured by means of a non-negative matrix factorization (NMF) model in \cite{ozerov2010multichannel,arberet2010nonnegative, ozerov2012general, sawada2013multichannel}. More recently, within this local Gaussian modeling framework based on SCMs, deep neural networks (DNNs) were investigated as a variance model \cite{nugraha_taslp16}. However, in \cite{nugraha_taslp16} the DNN is not really integrated in the generative source modeling, it is rather used as an ad-hoc ``deterministic'' variance estimator. 

For single-channel speech enhancement, DNN-based discriminative approaches have been successfully investigated for estimating either time-frequency masks or clean spectrograms from noisy spectrograms \cite{wang2017supervised}. Very recently, DNN-based generative models, and in particular variational autoencoders (VAEs) \cite{kingma2013auto}, have been employed for single-channel speech enhancement \cite{bando2017statistical,Leglaive_MLSP18,Leglaive_ICASSP2019b}. In \cite{Leglaive_MLSP18} we highlighted conceptual similarities between this approach and a more conventional generative model based on Itakura-Saito NMF \cite{ISNMF}. We also showed that the VAE-based speech model outperformed its NMF-based counterpart.

In this paper, we present an extension of \cite{Leglaive_MLSP18} to the multichannel case. The speech and noise models are based on the above-mentioned multichannel local Gaussian modeling approach. The speech source spectro-temporal content is further modeled in a supervised way by means of a DNN, whose parameters are learned using the VAE framework. We do not assume any prior knowledge on the recording environment, so the multichannel images of both the speech and the noise signals include a free SCM model. Moreover the noise source spectro-temporal model remains unsupervised and is based on NMF. We propose a Monte Carlo expectation-maximization (MCEM) algorithm \cite{wei1990monte} for estimating the model parameters. Experiments show that our approach outperforms its NMF counterpart \cite{sawada2013multichannel}, where the speech signal is modeled using supervised NMF. Note that a similar model for multichannel speech enhancement was published independently in \cite{BayesianMVAE}, after we submitted this paper. It is also combines VAE and NMF within a multichannel local Gaussian modeling framework, but it relies on a fully Bayesian approach.


\section{VAE-based speech source model}
\label{sec:speech_prior}

\begin{figure}[t]
\centering
\subfloat[Generative network.]{\includegraphics[width=.41\linewidth]{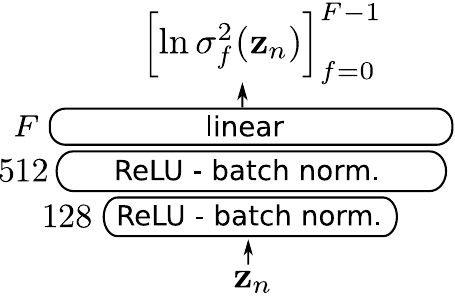}\label{fig:genNet}}
\hspace{0.2cm}
\subfloat[Recognition network.]{\includegraphics[width=.49\linewidth]{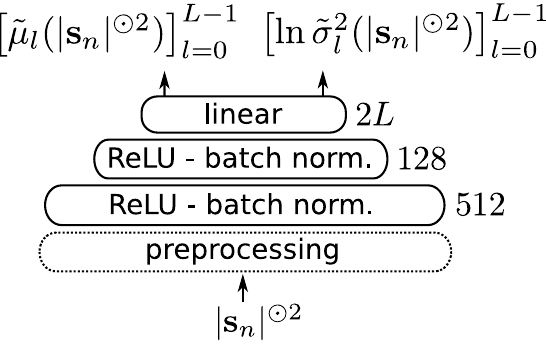}\label{fig:recNet}}
\caption{Neural network architectures. The number next to each layer indicates its size. Further architecture details are given in Section~\ref{sec:experiments}.}
\label{fig:fullVAE}
\end{figure}

We work in the short-term Fourier transform (STFT) domain where $\mbb{B}={\{0,...,F-1\}}\times{\{0,...,N-1\}}$ denotes the set of time-frequency bins. For  $(f,n) \in \mbb{B}$, $f$ denotes the frequency index and $n$ the time-frame index. As in \cite{bando2017statistical, Leglaive_MLSP18}, independently for every $(f,n) \in \mathbb{B}$, we consider the following generative speech source model involving a latent random vector $\mathbf{z}_n \in \mathbb{R}^L$:
\begin{align}
\mathbf{z}_n &\sim \mathcal{N}(\mathbf{0}, \mathbf{I}); \label{prior_VAE} \\
s_{fn} \mid \mathbf{z}_n &\sim \mathcal{N}_c(0, \sigma_f^2(\mathbf{z}_n)),
\label{decoder_VAE}
\end{align}
where $\mathcal{N}(\bs{\mu}, \bs{\Sigma})$ denotes the multivariate Gaussian distribution for a real-valued random vector, $\mathcal{N}_c(\mu, \sigma^2)$ denotes the univariate complex proper \cite{properComplex} Gaussian distribution and $\mathbf{I}$ is the identity matrix of appropriate size. As represented in Fig.~\ref{fig:genNet}, $\{\sigma_f^2: \mbb{R}^L \mapsto \mbb{R}_+\}_{f=0}^{F-1}$ is a set of non-linear functions provided by a DNN which takes as input $\mbf{z}_n \in \mathbb{R}^L$. In the following, we will denote by $\bs{\theta}_s$ the parameters of this \emph{generative network}, which can be seen as a model of the speech short-term power spectral density \cite{liutkus2011gaussian}.

A key contribution of VAEs is to provide an efficient way of learning the parameters $\bs{\theta}_s$ of such a generative model \cite{kingma2013auto}. Let $\mbf{s} = \{ \mbf{s}_n \in \mbb{C}^F \}_{n=0}^{N-1}$ be a training dataset of clean-speech STFT spectra and $\mbf{z} = \{ \mbf{z}_n \in \mbb{R}^L \}_{n=0}^{N-1}$ the set of associated latent random vectors. Taking ideas from variational inference, the parameters $\bs{\theta}_s$ are estimated by maximizing a lower bound $\mathcal{L}\left(\bs{\theta}_s, \bs{\phi}\right)$ of the log-likelihood $\ln p(\mbf{s} ; \bs{\theta}_s)$ defined by:
\begin{equation}
\mathcal{L}\left(\bs{\theta}_s, \bs{\phi}\right) =  \E_{q\left(\mathbf{z} \mid \mathbf{s} ; \bs{\phi}\right)}\left[ \ln p\left(\mathbf{s} \mid \mathbf{z} ; \bs{\theta}_s \right) \right] - D_{\text{KL}}\left(q\left(\mathbf{z} \mid \mathbf{s} ; \bs{\phi}\right) \parallel p(\mathbf{z})\right),
\label{variational_bound}
\end{equation}
where $q\left(\mathbf{z} \mid \mathbf{s} ; \bs{\phi}\right)$ approximates the intractable true posterior distribution $p(\mathbf{z} \mid \mathbf{s} ; \bs{\theta}_s )$, and $D_{\text{KL}}(q \parallel p) = \mathbb{E}_q[\ln(q/p)]$ is the Kullback-Leibler divergence. Independently for all the dimensions $l \in \{0,...,L-1\}$ and all the time frames $n \in \{0,...,N-1\}$, $q(\mathbf{z} \mid \mathbf{s} ; \bs{\phi})$ is defined by:
\begin{equation}
z_{l,n} \mid \mathbf{s}_n \sim \mathcal{N}\left(\tilde{\mu}_l\left(|\mathbf{s}_n|^{\odot 2}\right), \tilde{\sigma}_l^2\left(|\mathbf{s}_n|^{\odot 2}\right) \right),
\label{encoder_VAE}
\end{equation}
where $z_{l,n} = (\mathbf{z}_n)_l$ and $(\cdot)^{\odot \cdot}$ denotes element-wise exponentiation. As represented in Fig.~\ref{fig:recNet}, $\{\tilde{\mu}_l: \mathbb{R}_+^{F} \mapsto \mathbb{R}\}_{l=0}^{L-1}$ and $\{\tilde{\sigma}_l^2: \mathbb{R}_+^{F} \mapsto \mathbb{R}_+\}_{l=0}^{L-1}$ are non-linear functions provided by a DNN which takes as input the speech power spectrum at a given time frame. $\bs{\phi}$ denotes the parameters of this \emph{recognition network}, which also have to be estimated by maximizing the \emph{variational lower bound} defined in \eqref{variational_bound}. Using \eqref{prior_VAE}, \eqref{decoder_VAE} and \eqref{encoder_VAE} we can develop \eqref{variational_bound} as follows:
\begin{align}
\mathcal{L}\left(\bs{\theta}_s, \bs{\phi}\right) \overset{c}{=}&- \sum_{f=0}^{F-1}\sum_{n=0}^{N-1} \mathbb{E}_{q\left(\mathbf{z}_n \mid \mathbf{s}_n ; \bs{\phi} \right)}\left[ d_{\text{IS}}\left(\left|s_{fn}\right|^2 ; \sigma_f^2(\mathbf{z}_n)\right) \right] \nonumber \\
& \hspace{-1.6cm} + \frac{1}{2} \sum_{l=0}^{L-1} \sum_{n=0}^{N-1}\left[ \ln \tilde{\sigma}_l^2\left(\left|\mathbf{s}_n\right|^{\odot 2}\right) - \tilde{\mu}_l\left(\left|\mathbf{s}_n\right|^{\odot 2}\right)^2 - \tilde{\sigma}_l^2\left(\left|\mathbf{s}_n\right|^{\odot 2}\right) \right],
\label{ELBO}
\end{align}
where $\overset{c}{=}$ denotes equality up to an additive constant and $d_{\text{IS}}(x;y) = x/y - \ln(x/y) - 1$ is the Itakura-Saito divergence. Finally, using the so-called ``reparametrization trick'' \cite{kingma2013auto} to approximate the intractable expectation in \eqref{ELBO}, we obtain an objective function which is differentiable with respect to both $\bs{\theta}_s$ and $\bs{\phi}$ and can be optimized using gradient-ascent-based algorithms. It is important to note that the only reason why the recognition network is introduced is to learn the parameters of the generative network.

\section{Complete Multichannel Model}
\label{sec:model}
In the previous section we have seen how to learn the parameters of the generative model \eqref{prior_VAE}-\eqref{decoder_VAE}. This model can then be used as a speech source signal probabilistic prior for a variety of applications. In this paper we are interested in multichannel speech enhancement. We now present the model developed for this task. In the following, $I$ denotes the number of microphones, and probabilistic models are defined independently for all time-frequency points $(f,n) \in \mbb{B}$.

The multichannel speech signal $\mbf{s}_{fn} \in \mbb{C}^I$ is modeled as follows:
\begin{align}
\mbf{s}_{fn} \mid \mathbf{z}_n &\sim \mathcal{N}_c\big(\mbf{0}, \bs{\Sigma}_{\mbf{s},f}(\mathbf{z}_n) \big), 
\label{speech_model}
\end{align}
where $\bs{\Sigma}_{\mbf{s},f}(\mathbf{z}_n) \mydef \sigma_f^2(\mbf{z}_n) \mbf{R}_{\mbf{s},f}$. $\sigma_f^2(\cdot)$ was introduced in \eqref{decoder_VAE} and models the speech signal power spectral density, $\mathbf{z}_n$ follows the prior in \eqref{prior_VAE}, $\mbf{R}_{\mbf{s},f} \in \mbb{C}^{I \times I}$ is the speech signal spatial covariance matrix (which is Hermitian positive definite)  \cite{duong2010under} and $\mathcal{N}_c(\bs{\mu}, \bs{\Sigma})$ is the multivariate complex proper Gaussian distribution. 

The multichannel noise signal $\mbf{b}_{fn} \in \mbb{C}^I$ is modeled as follows:
\begin{equation}
\mbf{b}_{fn} \sim \mathcal{N}_c\big(\mbf{0}, \bs{\Sigma}_{\mbf{b},fn} \big),
\label{noise_model}
\end{equation}
where $\bs{\Sigma}_{\mbf{b},fn} \mydef \left(\mbf{W}_b\mbf{H}_b\right)_{f,n} \mbf{R}_{\mbf{b},f}$, with $\mathbf{W}_b = [w_{b,fk}]_{f,k} \in \mbb{R}_+^{F \times K_b}$, $\mathbf{H}_b = [h_{b,kn}]_{k,n} \in \mbb{R}_+^{K_b \times N}$ and $\mbf{R}_{\mbf{b},f} \in \mbb{C}^{I \times I}$ is the noise SCM (which is also Hermitian positive definite). The noise spectro-temporal content is thus modeled by NMF \cite{ISNMF}, and it remains unsupervised, i.e. $\mathbf{W}_b$ and $\mathbf{H}_b$ will be estimated from test data (and so will be the speech and noise SCMs).

The noisy mixture signal $\mbf{x}_{fn} \in \mbb{C}^I$ is finally modeled by:
\begin{equation}
\mbf{x}_{fn} = \sqrt{g_n} \mbf{s}_{fn} + \mbf{b}_{fn},
\end{equation}
where $g_n \in \mbb{R}_+$ is a gain parameter whose importance has been experimentally shown in \cite{Leglaive_MLSP18}. We further assume the independence of the speech and noise signals so that:
\begin{equation}
\mbf{x}_{fn} \mid \mathbf{z}_n \sim \mathcal{N}_c(\mbf{0}, \bs{\Sigma}_{\mbf{x},fn}(\mathbf{z}_n) ),
\label{likelihood}
\end{equation}
where $\bs{\Sigma}_{\mbf{x},fn}(\mathbf{z}_n) \mydef g_n\bs{\Sigma}_{\mbf{s},f}(\mathbf{z}_n) + \bs{\Sigma}_{\mbf{b},fn}$.

\section{MCEM algorithm and inference}
\label{sec:inference}

Let $\bs{\theta}_u = \left\{\mathbf{W}_b, \mathbf{H}_b, \left\{\mbf{R}_{\mbf{s},f}, \mbf{R}_{\mbf{b},f}\right\}_{f=0}^{F-1}, \mbf{g}=[g_0,...,g_{N-1}]^\top \right\}$ be the set of model parameters to be estimated. In this section we develop an MCEM algorithm \cite{wei1990monte} for this aim. Remember that the parameters $\bs{\theta}_s$ of the speech source model have been learned during a training phase (see Section~\ref{sec:speech_prior}). The set of observed data is denoted by $\mathbf{x} = \{\mbf{x}_{fn}\}_{(f,n) \in \mbb{B}}$ while the set of latent variables is $\mathbf{z} = \{\mbf{z}_{n}\}_{n=0}^{N-1}$. We will also use $\mathbf{x}_n = \{\mbf{x}_{fn}\}_{f=0}^{F-1}$ to denote all the observations for a given time frame $n \in \{0,...,N-1\}$.

\myparagraph{Monte Carlo E-Step} Let $\bs{\theta}_u^\star$ denote the current model parameters. At the E-step of a standard EM algorithm, we would compute the following conditional expectation of the complete-data log-likelihood $Q(\bs{\theta}_u ; \bs{\theta}_u^\star) = \mathbb{E}_{p(\mathbf{z} \mid \mathbf{x} ; \bs{\theta}_s, \bs{\theta}_u^\star)} \left[ \ln p(\mathbf{x}, \mathbf{z} ; \bs{\theta}_s, \bs{\theta}_u) \right]$. However, due to the non-linear relationship between the latent variables and the observations, we cannot compute this expectation in an analytical form. We thus approximate $Q(\bs{\theta}_u ; \bs{\theta}_u^\star)$ using an empirical average:
\begin{align}
\tilde{Q}(\bs{\theta}_u; \bs{\theta}_u^\star) \overset{c}{=} & - \frac{1}{R} \sum_{r=1}^{R} \sum_{(f,n) \in \mathbb{B}} \Big[ \trace\Big(\mbf{x}_{fn} \mbf{x}_{fn}^H \left[\bs{\Sigma}_{\mbf{x},fn}\big(\mathbf{z}_n^{(r)}\big)\right]^{-1} \Big) \nonumber \\ 
&  + \ln \det\Big( \bs{\Sigma}_{\mbf{x},fn}\big(\mathbf{z}_n^{(r)}\big)\Big) \Big],
\label{Q-function-tilde}
\end{align}
where $\cdot^H$ denotes the Hermitian transpose operator, $\trace(\cdot)$ and $\det(\cdot)$ denote respectively the trace and determinant of a matrix, and $\{\mathbf{z}_n^{(r)}\}_{r=1}^{R}$ is a sequence of samples asymptotically drawn from the posterior $p(\mathbf{z}_n\!\mid\!\mathbf{x}_n ; \bs{\theta}_s, \bs{\theta}_u^\star)$ using a Markov chain Monte Carlo (MCMC) method. This approach forms the basis of the MCEM algorithm. Here we use the Metropolis-Hastings algorithm \cite{Robert:2005:MCS:1051451}. Note that unlike the standard EM algorithm, the likelihood is not guaranteed to increase at each iteration. Nevertheless, some convergence results in terms of stationary point of the likelihood can be obtained under suitable conditions \cite{chan1995monte}. 

At the $m$-th iteration of the Metropolis-Hastings algorithm and independently for all $n \in \{0,...,N-1\}$, we first draw a sample $\tilde{\mbf{z}}_n$ from a random walk proposal distribution: $\mbf{z}_n \mid \mbf{z}_n^{(m-1)} \sim \mathcal{N}(\mbf{z}_n^{(m-1)}, \epsilon^2 \mathbf{I})$. Using the symmetry of this proposal distribution, we then compute the following acceptance probability:
\begin{align}
\alpha_n &= \min \Bigg(1, \frac{p(\tilde{\mbf{z}}_n) \prod\nolimits_{f=0}^{F-1} p(\mbf{x}_{fn} \mid \tilde{\mbf{z}}_n ; \bs{\theta}_s, \bs{\theta}_u^\star) }{p\left(\mbf{z}_n^{(m-1)}\right) \prod\nolimits_{f=0}^{F-1} p\left(\mbf{x}_{fn} \mid \mbf{z}_n^{(m-1)} ; \bs{\theta}_s, \bs{\theta}_u^\star\right) }\Bigg).
\end{align}
Then we draw $u_n$ from a uniform distribution $\mathcal{U}\left([0,1]\right)$. If $u_n < \alpha_n$, we accept the sample and set $\mbf{z}_n^{(m)} = \tilde{\mbf{z}}_n$, otherwise we reject the sample and set $\mbf{z}_n^{(m)} = \mbf{z}_n^{(m-1)}$. We only keep the last $R$ samples for computing $\tilde{Q}(\bs{\theta}_u; \bs{\theta}_u^\star)$, i.e. we discard the samples drawn during a so-called burn-in period.

\myparagraph{M-Step} At the M-step, we want to minimize the cost function $ \mathcal{C}(\bs{\theta}_u) \mydef -RQ(\bs{\theta}_u; \bs{\theta}_u^\star)$ with respect to $\bs{\theta}_u$. Similarly to \cite{sawada2013multichannel}, we adopt a majorization-minimization approach for solving this optimization problem. From \eqref{Q-function-tilde}, \eqref{likelihood}, \eqref{noise_model}, \eqref{speech_model} and using inequalities defined in Appendix~\ref{appendix:inequalities}, we can show that $\mathcal{C}(\bs{\theta}_u) \le \mathcal{G}(\bs{\theta}_u, \bs{\Phi}, \bs{\Omega})$ where the majorizing function is defined by:
\begin{align}
&\mathcal{G}(\bs{\theta}_u, \bs{\Phi}, \bs{\Omega}) = - IFNR \nonumber \\
&+ \sum_{r=1}^{R} \sum_{(f,n) \in \mathbb{B}} \Bigg[ \frac{1}{g_n \sigma_f^2\left(\mathbf{z}_n^{(r)}\right)} \trace\left(\mbf{x}_{fn} \mbf{x}_{fn}^H \left(\bs{\Phi}_{0,fn}^{(r)}\right)^H \mbf{R}_{\mbf{s},f}^{-1} \bs{\Phi}_{0,fn}^{(r)} \right) \nonumber \\
& \hspace{.75cm} + \sum_{k=1}^{K_b} \Bigg( \frac{1}{w_{b,fk}h_{b,kn}} \trace\left(\mbf{x}_{fn} \mbf{x}_{fn}^H \left(\bs{\Phi}_{k,fn}^{(r)}\right)^H \mbf{R}_{\mbf{b},f}^{-1} \bs{\Phi}_{k,fn}^{(r)} \right) \nonumber \\
& \hspace{2cm} + w_{b,fk}h_{b,kn} \trace\left( \left(\bs{\Omega}_{fn}^{(r)}\right)^{-1} \mbf{R}_{\mbf{b},f} \right) \Bigg) \nonumber \\
& \hspace{.75cm} + g_n \sigma_f^2\left(\mathbf{z}_n^{(r)}\right) \trace\left( \left(\bs{\Omega}_{fn}^{(r)}\right)^{-1} \mbf{R}_{\mbf{s},f} \right) + \ln \det\left(\bs{\Omega}_{fn}^{(r)}\right)\Bigg],
\label{upper_bound}
\end{align}
where $\bs{\Phi} = \left\{\bs{\Phi}_{k,fn}^{(r)} \in \mbb{C}^{I \times I}\right\}_{r,k,f,n}$, $\bs{\Omega} = \left\{\bs{\Omega}_{fn}^{(r)} \in \mbb{C}^{I \times I} \right\}_{r,f,n}$ are \emph{auxiliary variables}. This upper bound is tight, i.e. $\mathcal{C}(\bs{\theta}_u) = \mathcal{G}(\bs{\theta}_u, \bs{\Phi}, \bs{\Omega})$ if
\begin{flalign}
\bs{\Omega}_{fn}^{(r)} &= \bs{\Sigma}_{\mbf{x},fn}\left(\mathbf{z}_n^{(r)}\right);& \label{omega}\\
\bs{\Phi}_{0,fn}^{(r)} &= g_n \sigma_f^2\left(\mathbf{z}_n^{(r)}\right) \mbf{R}_{\mbf{s},f} \left[\bs{\Sigma}_{\mbf{x},fn}\left(\mathbf{z}_n^{(r)}\right)\right]^{-1};& \label{phi_0}\\
\bs{\Phi}_{k,fn}^{(r)} &= w_{b,fk}h_{b,kn} \mbf{R}_{\mbf{b},f} \left[\bs{\Sigma}_{\mbf{x},fn}\left(\mathbf{z}_n^{(r)}\right)\right]^{-1} \forall k \in \{1,...,K_b\}.&\label{phi_k}
\end{flalign}
Although not jointly convex, $\mathcal{G}(\bs{\theta}_u, \bs{\Phi}, \bs{\Omega})$ is separately convex in each of the individual parameters $g_n$, $w_{b,fk}$, $h_{b,kn}$, $\mbf{R}_{\mbf{s},f}$ and $\mbf{R}_{\mbf{b},f}$. We will therefore update individually and alternatively those parameters, which corresponds to a (block) coordinate approach. For one of those parameters, the general procedure consists in canceling the partial derivative of $\mathcal{G}(\bs{\theta}_u, \bs{\Phi}, \bs{\Omega})$. In this way we obtain an update that depends on the auxiliary variables. We then inject in this update the expressions of the auxiliary variables that make the upper bound tight, which are given by \eqref{omega}-\eqref{phi_k}. For notational convenience, let us first introduce:
\begin{equation}
\mbf{M}_{fn}^{(r)} := \left(\bs{\Sigma}_{\mbf{x},fn}\left(\mathbf{z}_n^{(r)}\right)\right)^{-1} \mbf{x}_{fn} \mbf{x}_{fn}^H \left(\bs{\Sigma}_{\mbf{x},fn}\left(\mathbf{z}_n^{(r)}\right)\right)^{-1}.
\end{equation}
The resulting updates are given as follows:
\begin{equation}
w_{fk} \leftarrow w_{fk} \left[\frac{\sum\limits_{r=1}^{R} \sum\limits_{n=0}^{N-1} h_{kn} \trace\left[ \mbf{M}_{fn}^{(r)} \mbf{R}_{\mbf{b},f} \right] }{\sum\limits_{r=1}^{R} \sum\limits_{n=0}^{N-1} h_{kn} \trace\left[ \left(\bs{\Sigma}_{\mbf{x},fn}\left(\mathbf{z}_n^{(r)}\right)\right)^{-1} \mbf{R}_{\mbf{b},f} \right] } \right]^{1/2} ;
\label{upd_w}
\end{equation}
\begin{equation}
h_{kn} \leftarrow h_{kn} \left[\frac{\sum\limits_{r=1}^{R} \sum\limits_{f=0}^{F-1} w_{fk} \trace\left[ \mbf{M}_{fn}^{(r)} \mbf{R}_{\mbf{b},f} \right] }{\sum\limits_{r=1}^{R} \sum\limits_{f=0}^{F-1} w_{fk} \trace\left[ \left(\bs{\Sigma}_{\mbf{x},fn}\left(\mathbf{z}_n^{(r)}\right)\right)^{-1} \mbf{R}_{\mbf{b},f} \right] } \right]^{1/2} ;
\label{upd_h}
\end{equation}
\begin{equation}
\hspace{-0.1cm}g_{n} \leftarrow g_{n} \left[\frac{\sum\limits_{r=1}^{R} \sum\limits_{f=0}^{F-1} \sigma_f^2\left(\mathbf{z}_n^{(r)}\right) \trace\left[\mbf{M}_{fn}^{(r)} \mbf{R}_{\mbf{s},f} \right] }{\sum\limits_{r=1}^{R} \sum\limits_{f=0}^{F-1} \sigma_f^2\left(\mathbf{z}_n^{(r)}\right) \trace\left[ \left(\bs{\Sigma}_{\mbf{x},fn}\left(\mathbf{z}_n^{(r)}\right)\right)^{-1} \mbf{R}_{\mbf{s},f} \right] } \right]^{1/2}\hspace{-0.1cm}.
\label{upd_g}
\end{equation}
Updating the SCMs is done by solving the two following algebraic Riccati equations (see the procedure in \cite[Appendix I]{sawada2013multichannel}):
\begin{align}
\mbf{R}_{\mbf{s},f} & \left[\sum\limits_{r=1}^{R} \sum\limits_{n=0}^{N-1} g_n \sigma_f^2\left(\mathbf{z}_n^{(r)}\right) \left(\bs{\Sigma}_{\mbf{x},fn}\left(\mathbf{z}_n^{(r)}\right)\right)^{-1} \right] \mbf{R}_{\mbf{s},f} \nonumber \\
& = \tilde{\mbf{R}}_{\mbf{s},f} \left[\sum\limits_{r=1}^{R} \sum\limits_{n=0}^{N-1} g_n \sigma_f^2\left(\mathbf{z}_n^{(r)}\right) \mbf{M}_{fn}^{(r)}  \right] \tilde{\mbf{R}}_{\mbf{s},f};
\label{upd_R_s}
\end{align}
\begin{align}
\mbf{R}_{\mbf{b},f} & \left[\sum\limits_{r=1}^{R} \sum\limits_{n=0}^{N-1} \left(\mbf{W}_b\mbf{H}_b\right)_{f,n} \left(\bs{\Sigma}_{\mbf{x},fn}\left(\mathbf{z}_n^{(r)}\right)\right)^{-1} \right] \mbf{R}_{\mbf{b},f} \nonumber \\
& = \tilde{\mbf{R}}_{\mbf{b},f} \left[\sum\limits_{r=1}^{R} \sum\limits_{n=0}^{N-1} \left(\mbf{W}_b\mbf{H}_b\right)_{f,n} \mbf{M}_{fn}^{(r)}  \right] \tilde{\mbf{R}}_{\mbf{b},f},
\label{upd_R_b}
\end{align}
where $\tilde{\mbf{R}}_{\mbf{s},f}$ and $\tilde{\mbf{R}}_{\mbf{b},f}$ denote the SCMs before update. Note that $\bs{\Sigma}_{\mbf{x},fn}\left(\mathbf{z}_n^{(r)}\right)$ and therefore $\mbf{M}_{fn}^{(r)}$ should be updated after each one of the updates \eqref{upd_w}-\eqref{upd_R_b}.

The criterion $\mathcal{C}(\bs{\theta}_u)$ suffers from scale indeterminacies between $\mbf{R}_{\mbf{b},f}$ and $w_{b,fk}$, and also between $w_{b,fk}$ and $h_{b,kn}$. At each iteration of the algorithm, we compensate these indeterminacies by normalizing the noise SCM so that $\trace(\mbf{R}_{\mbf{b},f}) = 1$ (and scaling the rows of $\mbf{W}_b$ accordingly so that the criterion remains unchanged) and by normalizing the noise dictionary matrix so that $\sum_{f=0}^{F-1} w_{b,fk} = 1$ (and scaling the columns of $\mbf{H}_b$ accordingly).

\myparagraph{Speech reconstruction} Once the unsupervised model parameters are estimated with the MCEM algorithm, we need to estimate the multichannel speech signal. For all $(f,n) \in \mbb{B}$, let $\tilde{\mbf{s}}_{fn} = \sqrt{g_n} \mbf{s}_{fn}$ be the scaled version of the multichannel speech signal. 
We estimate this vector according to its posterior mean:
\begin{align}
\hat{\tilde{\mbf{s}}}_{fn} &= \mbb{E}_{p(\tilde{\mbf{s}}_{fn} \mid \mbf{x}_{fn} ; \bs{\theta}_s, \bs{\theta}_u)} \left[ \tilde{\mbf{s}}_{fn} \right] \nonumber \\
&= \mbb{E}_{p(\mbf{z}_{n} \mid \mbf{x}_{n} ; \bs{\theta}_s, \bs{\theta}_u)} \left[ \mbb{E}_{p(\tilde{\mbf{s}}_{fn} \mid \mbf{x}_{n}, \mbf{z}_{n} ; \bs{\theta}_s, \bs{\theta}_u)} \left[ \tilde{\mbf{s}}_{fn} \right]  \right] \nonumber \\
&= \mbb{E}_{p(\mbf{z}_{n} \mid \mbf{x}_{n} ; \bs{\theta}_s, \bs{\theta}_u)} \left[ g_n \bs{\Sigma}_{\mbf{s},f}(\mathbf{z}_n) \left(\bs{\Sigma}_{\mbf{x},fn}\left(\mathbf{z}_n\right)\right)^{-1} \right]\mbf{x}_{fn}.
\label{eq:speech-estimate}
\end{align}
As before, \eqref{eq:speech-estimate} cannot be computed in an analytical form, and it is approximated with the  Metropolis-Hastings algorithm as done in the E-Step. 
This speech signal estimate corresponds to a probabilistic multichannel Wiener filtering averaged over all possible realizations of the latent variables according to their posterior distribution.
Note that the noise signal can be estimated in a similar way.

\section{Experiments}
\label{sec:experiments}

\begin{figure*}[!ht]
\centering
\includegraphics[width=.9\linewidth]{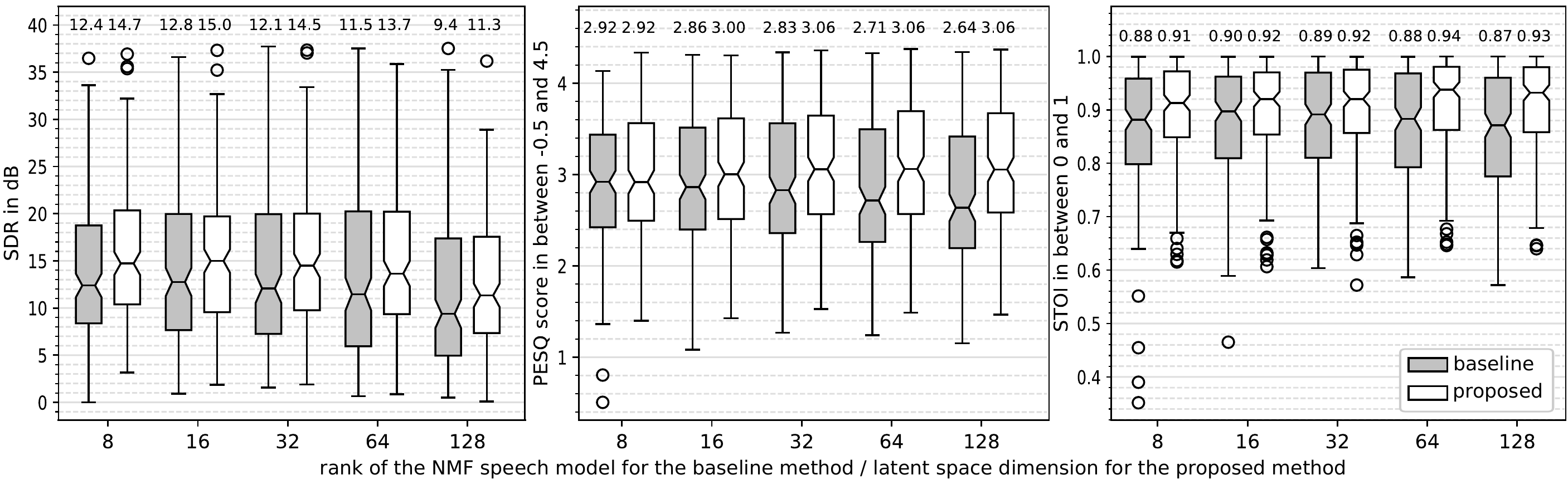}
\caption{Speech enhancement results in terms of SDR, PESQ and STOI measures. The value of the median is indicated above each boxplot.}
\label{fig:results}
\vspace{-0.4cm}
\end{figure*}

\textit{Baseline method}: We compare our method with its NMF counterpart proposed in \cite{sawada2013multichannel}. In this paper, the multichannel speech signal is modeled by $\mbf{s}_{fn} \sim \mathcal{N}_c(0, \left(\mbf{W}_s\mbf{H}_s\right)_{f,n} \mbf{R}_{\mbf{s},f} ),$ where $\mathbf{W}_s \in \mbb{R}_+^{F \times K_s}$ and $\mathbf{H}_b \in \mbb{R}_+^{K_s \times N}$. Compared with the proposed model defined in \eqref{speech_model}, the deep-learning based variance model is replaced by a more conventional NMF model. For this baseline method, we also consider a supervised speech model by learning the NMF dictionary matrix $\mbf{W}_s$ on the training dataset. The noise model remains unsupervised and is identical to the one defined in \eqref{noise_model}.

\textit{Database}: The supervised speech model parameters are learned from the training set of the TIMIT database \cite{TIMIT}. It contains almost 4 hours of single-channel 16-kHz speech signals, distributed over 462 speakers. For the evaluation of the speech enhancement algorithms, we created 168 stereophonic noisy mixtures of about 3 seconds each. We used the monophonic clean speech signals from the TIMIT test set (speakers and sentences are different than in the training set) and converted them to stereophonic signals as follows: We uniformly drew a direction of arrival within [-90, 90]° and we delayed one channel with respect to the other accordingly, assuming a free field propagation. We then mixed the resulting signals with stereophonic non-stationary noise signals from the DEMAND database \cite{DEMAND_db} (using channels 1 and 3), at a 0 dB signal-to-noise ratio. The different noisy recording environments are domestic, nature, office, indoor public spaces, street and transportation. 

\textit{Parameters setting}: The STFT is computed using a 64-ms sine window (i.e.~$F=513$) with 75\%-overlap. The proposed and baseline methods are compared for different settings of the dimension of the latent space $L$ and the rank $K_s$ of the NMF speech model. For fair comparison, we set $L = K_s \in \{8, 16, 32, 64, 128\}$. The rank of the noise model is arbitrarily fixed to $K_b = 10$ for both methods. Unsupervised NMF parameters are randomly initialized and SCMs are initialized with identity matrices. For the proposed method, $\mathbf{g}$ is initialized with an all-ones vector. Similarly as in \cite{Leglaive_MLSP18}, the Markov chain at the first MCEM iteration is initialized using the recognition network taking the noisy mixture as input.  Then, at each new E-Step, we use the last sample drawn at the previous E-Step to initialize the Markov chain. We then run 40 iterations of the Metropolis-Hastings algorithm with $\epsilon^2 = 0.01$, and we discard the first 30 samples as the burn-in period. The algorithms for the proposed and the baseline methods are run for 50 iterations.

\textit{Neural network}: The generative and recognition networks are represented in Fig.~\ref{fig:fullVAE}. Each network is feed-forward and fully-connected, with hidden layers using rectified linear units (ReLUs) \cite{relu} and batch normalization \cite{batch_norm}. Output layers use identity activation functions, so they output real-valued coefficients which is the reason why we consider logarithm of variances. The preprocessing layer of the recognition network consists in computing the logarithm of the input speech power spectra and applying a frequency-wise standardization (zero-mean and unit-variance) based on the training set. For jointly learning the parameters of the generative and recognition networks (see Section~\ref{sec:speech_prior}), we use the Adam optimizer \cite{kingma2014adam} with a step size of $10^{-3}$, exponential decay rates of $0.9$ and $0.999$ for the first and second moment estimates, respectively, and an epsilon of $10^{-7}$ for preventing division by zero. 20\% of the TIMIT training set was kept as a validation set, and early stopping with a patience of 10 epochs was used. Weights were initialized using the uniform initializer described in \cite{glorot2010understanding}. The network architecture was chosen based on the cost function on the validation set.

\textit{Results}: The speech enhancement results are presented in Fig.~\ref{fig:results} in terms of signal-to-distortion ratio (SDR) \cite{vincent2006performance}, perceptual evaluation of speech quality (PESQ) measure \cite{rix2001perceptual} and short-time objective intelligibility (STOI) measure \cite{taal2011algorithm}. We see that according to all measures, the proposed method outperforms its NMF-based counterpart. In terms of SDR, for the best configuration of the two methods ($L = K_s = 16$), we have a 2.2 dB improvement in median. It is also interesting to observe that all the performance measures for the baseline method tend to decrease as $K_s$ increases (above $16$). Increasing the number of parameters makes the optimization problem more difficult to solve. For the proposed method, this behaviour also appears (although to a lesser extent) in terms of SDR only, not in terms PESQ or STOI. Our better performance is however obtained at the expense of a higher computational time. Due to the Monte Carlo E-Step, and with the above-mentioned parameters setting, one iteration of the proposed method is about seven times more expensive than one iteration of the baseline method. For reproducibility, a Python implementation of our algorithm and audio examples are available online.\footnote{\url{https://team.inria.fr/perception/icassp-2019-mvae/}}
\vspace{-0.2cm}

\section{Conclusion}
\label{sec:conclusion}

In this work, we proposed a multichannel speech enhancement method using a probabilistic speech source model based on variational autoencoders. The noise model was based on unsupervised NMF, which makes the method nicely flexible and able to adapt to any (non-stationary) noisy environment. Experiments demonstrated that the proposed approach outperforms its NMF-based counterpart. The main limitation of this work is the increased computational time. In future work, we will replace the MCEM algorithm by a variational EM algorithm, in order to speed up the inference.
\vspace{-0.2cm}

\appendix
\section{Inequalities}
\label{appendix:inequalities}
\vspace{-0.2cm}

Let $\bs{\Sigma} \in \mathbb{C}^{I \times I}$ be a positive definite matrix. As $\bs{\Sigma} \mapsto \ln \det\left( \bs{\Sigma} \right)$ is a concave function, it can be majorized at an arbitrary point using a first-order Taylor expansion: $\ln \det\left( \bs{\Sigma} \right) \le \ln \det\left( \bs{\Sigma}_0 \right) + \trace\left( \bs{\Sigma}_0^{-1} \bs{\Sigma} \right) - I$, where $\bs{\Sigma}_0 \in \mathbb{C}^{I \times I}$ is a positive definite matrix \cite[p. 641]{boyd2004convex}. Moreover, equality holds if $\bs{\Sigma}_0 = \bs{\Sigma}$. 

For any positive definite matrix $\bs{\Sigma} \in \mathbb{C}^{I \times I}$, and any matrix $\mbf{A} \in \mathbb{C}^{I \times I}$, $\bs{\Sigma} \mapsto \trace\left( \bs{\Sigma}^{-1} \mbf{A} \right) $ is a convex function. Therefore, for any set of Hermitian positive definite matrices $\left\{ \bs{\Sigma}_k \in \mathbb{C}^{I \times I} \right\}_k$, Jensen's trace inequality gives $\trace\left( \left( \sum_{k} \bs{\Sigma}_k \right)^{-1} \mbf{A} \right) \le \sum_{k} \trace\left(\bs{\Sigma}_k^{-1} \bs{\Phi}_k \mbf{A} \bs{\Phi}_k^H \right)$, where $\bs{\Phi}_k \in \mathbb{C}^{I \times I}$ and $\sum_{k} \bs{\Phi}_k = \mbf{I}$ \cite{hansen2003jensen}. Moreover, equality holds if $\bs{\Phi}_k = \bs{\Sigma}_k \left(\sum_{k'} \bs{\Sigma}_{k'}\right)^{-1}$.


\unappendix
\balance
\bibliographystyle{IEEEbib}
\bibliography{IEEEabrv,refs}

\begin{thebibliography}{10}

\bibitem{loizou2007speech}
Philipos~C. Loizou,
\newblock {\em Speech enhancement: theory and practice},
\newblock CRC press, 2007.

\bibitem{ozerov2010multichannel}
Alexey Ozerov and C{\'e}dric F{\'e}votte,
\newblock ``Multichannel nonnegative matrix factorization in convolutive
  mixtures for audio source separation,''
\newblock {\em {IEEE} Trans. Audio, Speech, Language Process.}, vol. 18, no. 3,
  pp. 550--563, 2010.

\bibitem{duong2010under}
Ngoc Q.~K. Duong, Emmanuel Vincent, and R{\'e}mi Gribonval,
\newblock ``Under-determined reverberant audio source separation using a
  full-rank spatial covariance model,''
\newblock {\em {IEEE} Trans. Audio, Speech, Language Process.}, vol. 18, no. 7,
  pp. 1830--1840, 2010.

\bibitem{arberet2010nonnegative}
Simon Arberet, Alexey Ozerov, Ngoc~QK Duong, Emmanuel Vincent, R{\'e}mi
  Gribonval, Fr{\'e}d{\'e}ric Bimbot, and Pierre Vandergheynst,
\newblock ``Nonnegative matrix factorization and spatial covariance model for
  under-determined reverberant audio source separation,''
\newblock in {\em Proc. IEEE Int. Conf. Information Sciences, Signal Process.
  and Applications (ISSPA)}, 2010, pp. 1--4.

\bibitem{ozerov2012general}
Alexey Ozerov, Emmanuel Vincent, and Fr{\'e}d{\'e}ric Bimbot,
\newblock ``A general flexible framework for the handling of prior information
  in audio source separation,''
\newblock {\em {IEEE} Trans. Audio, Speech, Language Process.}, vol. 20, no. 4,
  pp. 1118--1133, 2012.

\bibitem{sawada2013multichannel}
Hiroshi Sawada, Hirokazu Kameoka, Shoko Araki, and Naonori Ueda,
\newblock ``Multichannel extensions of non-negative matrix factorization with
  complex-valued data,''
\newblock {\em {IEEE} Trans. Audio, Speech, Language Process.}, vol. 21, no. 5,
  pp. 971--982, 2013.

\bibitem{kitamura2016determined}
Daichi Kitamura, Nobutaka Ono, Hiroshi Sawada, Hirokazu Kameoka, and Hiroshi
  Saruwatari,
\newblock ``Determined blind source separation unifying independent vector
  analysis and nonnegative matrix factorization,''
\newblock {\em {IEEE} Trans. Audio, Speech, Language Process.}, vol. 24, no. 9,
  pp. 1626--1641, 2016.

\bibitem{nugraha_taslp16}
Aditya~Arie Nugraha, Antoine Liutkus, and Emmanuel Vincent,
\newblock ``{Multichannel audio source separation with deep neural networks},''
\newblock {\em {IEEE} Trans. Audio, Speech, Language Process.}, vol. 24, no. 9,
  pp. 1652--1664, 2016.

\bibitem{leglaive:taslp16}
Simon Leglaive, Roland Badeau, and Ga{\"e}l Richard,
\newblock ``Multichannel audio source separation with probabilistic
  reverberation priors,''
\newblock {\em {IEEE} Trans. Audio, Speech, Language Process.}, vol. 24, no.
  12, pp. 2453--2465, 2016.

\bibitem{wang2017supervised}
DeLiang Wang and Jitong Chen,
\newblock ``Supervised speech separation based on deep learning: An overview,''
\newblock {\em {IEEE} Trans. Audio, Speech, Language Process.}, vol. 26, no.
  10, pp. 1702--1726, 2018.

\bibitem{kingma2013auto}
Diederik~P. Kingma and Max Welling,
\newblock ``Auto-encoding variational {Bayes},''
\newblock in {\em Proc. Int. Conf. Learning Representations (ICLR)}, 2014.

\bibitem{bando2017statistical}
Yoshiaki Bando, Masato Mimura, Katsutoshi Itoyama, Kazuyoshi Yoshii, and
  Tatsuya Kawahara,
\newblock ``Statistical speech enhancement based on probabilistic integration
  of variational autoencoder and non-negative matrix factorization,''
\newblock in {\em Proc. IEEE Int. Conf. Acoust., Speech, Signal Process.
  (ICASSP)}, 2018, pp. 716--720.

\bibitem{Leglaive_MLSP18}
Simon Leglaive, Laurent Girin, and Radu Horaud,
\newblock ``A variance modeling framework based on variational autoencoders for
  speech enhancement,''
\newblock Proc. IEEE Int. Workshop Machine Learning Signal Process. (MLSP),
  2018.

\bibitem{Leglaive_ICASSP2019b}
Simon Leglaive, Umut \c{S}im\c{s}ekli, Antoine Liutkus, Laurent Girin, and Radu
  Horaud,
\newblock ``Speech enhancement with variational autoencoders and alpha-stable
  distributions,''
\newblock in {\em Proc. IEEE Int. Conf. Acoust., Speech, Signal Process.
  (ICASSP)}, 2019.

\bibitem{ISNMF}
C{\'e}dric F{\'e}votte, Nancy Bertin, and Jean-Louis Durrieu,
\newblock ``{Nonnegative matrix factorization with the Itakura-Saito
  divergence: With application to music analysis},''
\newblock {\em Neural computation}, vol. 21, no. 3, pp. 793--830, 2009.

\bibitem{wei1990monte}
Greg~C.G. Wei and Martin~A. Tanner,
\newblock ``A {Monte Carlo} implementation of the {EM} algorithm and the poor
  man's data augmentation algorithms,''
\newblock {\em Journal of the American statistical Association}, vol. 85, no.
  411, pp. 699--704, 1990.

\bibitem{BayesianMVAE}
Kouhei Sekiguchi, Yoshiaki Bando, Kazuyoshi Yoshii, and Tatsuya Kawahara,
\newblock ``Bayesian multichannel speech enhancement with a deep speech
  prior,''
\newblock in {\em Proc. Asia-Pacific Signal and Information Processing
  Association Annual Summit and Conference (APSIPA ASC)}, 2018, pp. 1233--1239.

\bibitem{properComplex}
F.~D. Neeser and J.~L. Massey,
\newblock ``Proper complex random processes with applications to information
  theory,''
\newblock {\em IEEE Trans. Information Theory}, vol. 39, no. 4, pp. 1293--1302,
  1993.

\bibitem{liutkus2011gaussian}
Antoine Liutkus, Roland Badeau, and G{\"a}el Richard,
\newblock ``Gaussian processes for underdetermined source separation,''
\newblock {\em {IEEE} Trans. Signal Process.}, vol. 59, no. 7, pp. 3155--3167,
  2011.

\bibitem{Robert:2005:MCS:1051451}
Christian~P. Robert and George Casella,
\newblock {\em {Monte Carlo} Statistical Methods},
\newblock Springer-Verlag New York, Inc., Secaucus, NJ, USA, 2005.

\bibitem{chan1995monte}
K.S. Chan and Johannes Ledolter,
\newblock ``{Monte Carlo EM} estimation for time series models involving
  counts,''
\newblock {\em Journal of the American Statistical Association}, vol. 90, no.
  429, pp. 242--252, 1995.

\bibitem{TIMIT}
John~S. Garofolo, Lori~F. Lamel, William~M. Fisher, Jonathan~G. Fiscus,
  David~S. Pallett, Nancy~L. Dahlgren, and Victor Zue,
\newblock ``{TIMIT} acoustic phonetic continuous speech corpus,''
\newblock in {\em Linguistic data consortium}, 1993.

\bibitem{DEMAND_db}
Joachim Thiemann, Nobutaka Ito, and Emmanuel Vincent,
\newblock ``{The Diverse Environments Multi-channel Acoustic Noise Database
  (DEMAND): A database of multichannel environmental noise recordings},''
\newblock in {\em {Proc. Int. Cong. on Acoust.}}, 2013.

\bibitem{relu}
Vinod Nair and Geoffrey~E. Hinton,
\newblock ``Rectified linear units improve restricted boltzmann machines,''
\newblock in {\em Proc. Int. Conf. Machine Learning (ICML)}, 2010, pp.
  807--814.

\bibitem{batch_norm}
Sergey Ioffe and Christian Szegedy,
\newblock ``{Batch Normalization: Accelerating Deep Network Training by
  Reducing Internal Covariate Shift},''
\newblock in {\em Proc. Int. Conf. Machine Learning (ICML)}, 2015, pp.
  448--456.

\bibitem{kingma2014adam}
Diederik~P. Kingma and Jimmy Ba,
\newblock ``Adam: A method for stochastic optimization,''
\newblock in {\em Proc. Int. Conf. Learning Representations (ICLR)}, 2015.

\bibitem{glorot2010understanding}
Xavier Glorot and Yoshua Bengio,
\newblock ``Understanding the difficulty of training deep feedforward neural
  networks,''
\newblock in {\em Proc. Int. Conf. Artif. Intelligence and Stat.}, 2010, pp.
  249--256.

\bibitem{vincent2006performance}
Emmanuel Vincent, R{\'e}mi Gribonval, and C{\'e}dric F{\'e}votte,
\newblock ``Performance measurement in blind audio source separation,''
\newblock {\em {IEEE} Trans. Audio, Speech, Language Process.}, vol. 14, no. 4,
  pp. 1462--1469, 2006.

\bibitem{rix2001perceptual}
Antony~W. Rix, John~G. Beerends, Michael~P. Hollier, and Andries~P. Hekstra,
\newblock ``{Perceptual evaluation of speech quality (PESQ)-a new method for
  speech quality assessment of telephone networks and codecs},''
\newblock in {\em Proc. IEEE Int. Conf. Acoust., Speech, Signal Process.
  (ICASSP)}, 2001, pp. 749--752.

\bibitem{taal2011algorithm}
Cees~H. Taal, Richard~C. Hendriks, Richard Heusdens, and Jesper Jensen,
\newblock ``An algorithm for intelligibility prediction of time--frequency
  weighted noisy speech,''
\newblock {\em {IEEE} Trans. Audio, Speech, Language Process.}, vol. 19, no. 7,
  pp. 2125--2136, 2011.

\bibitem{boyd2004convex}
Stephen Boyd and Lieven Vandenberghe,
\newblock {\em Convex optimization},
\newblock Cambridge university press, 2004.

\bibitem{hansen2003jensen}
Frank Hansen and Gert~K. Pedersen,
\newblock ``Jensen's operator inequality,''
\newblock {\em Bulletin of the London Mathematical Society}, vol. 35, no. 4,
  pp. 553--564, 2003.

\end{thebibliography}

\end{document}